\documentclass[twocolumn,showpacs,superscriptaddress,prl]{revtex4}
\usepackage{amsmath,amssymb}
\usepackage{graphicx}

\begin{document}

\title{Non-Abelian gauge potentials for ultra-cold atoms with
degenerate dark states}
\date{\today}
\author{J. Ruseckas}
\affiliation{Vilnius University Research Institute of Theoretical Physics and
Astronomy,\\
A. Go\v{s}tauto 12, 01108 Vilnius, Lithuania}
\affiliation{Fachbereich Physik, Technische Universit\"at Kaiserslautern,\\
D-67663 Kaiserslautern, Germany}
\author{G. Juzeli\={u}nas}
\affiliation{Vilnius University Research Institute of Theoretical Physics and
Astronomy,\\
A. Go\v{s}tauto 12, 01108 Vilnius, Lithuania}
\author{P. \"Ohberg}
\affiliation{Department of Physics, University of Strathclyde,\\
Glasgow G4 0NG, Scotland}
\author{M. Fleischhauer}
\affiliation{Fachbereich Physik, Technische Universit\"at Kaiserslautern,\\
D-67663 Kaiserslautern, Germany}
\begin{abstract}
We show that the adiabatic motion of ultra-cold, multi-level atoms in spatially
varying laser fields can give rise to
effective non-Abelian gauge fields if degenerate adiabatic 
eigenstates of the atom-laser interaction exist.
A pair of such degenerate dark states emerges
e.g. if laser fields couple three internal states of an atom to a fourth common
one under pairwise two--photon-resonance conditions. For this so-called tripod
scheme we derive general conditions for truly non-Abelian gauge potentials and
discuss special examples. In
particular we show that using orthogonal laser beams with orbital angular
momentum an effective magnetic field can be generated that has a monopole component.
\end{abstract}
\maketitle

Gauge fields are a central building block of the
theory of fundamental interactions. As dynamical variables they are responsible
for the forces between elementary particles.  On the other hand also
non-dynamical, i.e. prescribed gauge fields are of interest in a
variety of single- and many-body quantum systems.  E.g. an external magnetic
field applied to a gas of non-interacting electrons can lead to the integer
quantum Hall effect \cite{quantum-Hall}.  In the presence of a lattice
potential, the eigenenergies of the lowest Bloch band form a fractal structure
depending on the magnetic flux that passes through the unit cell
\cite{Hofstadter-PRB-1976}.  If in addition 
there are strong interactions between the
particles as e.g. in
a two dimensional electron gas subject to a magnetic field,
fractional quantum Hall structures \cite{FQHE} and Laughlin 
liquids \cite{Laughlin} can emerge.

In recent years ultra-cold atomic gases \cite{BEC} have become an
ideal playground to experimentally investigate many-body physics. This 
is due to their enormous versatility and the advanced experimental techniques 
available in atomic and optical physics. One of the
most fascinating subjects in this context is the study of effects of artificial
magnetic fields \cite{Joliceour}. To create an artificial magnetic field for
neutral atoms one can e.g. rotate the trapping potential confining
the atoms.  This experimentally feasible but challenging approach is currently
pursued in several labs \cite{rotating-traps}. An alternative
is based on the adiabatic motion of $\Lambda$-type 3-level atoms in laser
fields that create a non-degenerate dark state, i.e. an eigenstate of the atom-laser interaction.
If the dark-state of the atom is space dependent,
the motion of atoms adiabatically following it is associated with a topological
or Berry phase \cite{Berry-ProcRoySoc-1984,Shapere-book}.  A proper description
of such a motion naturally leads to Abelian gauge potentials
\cite{Jackiw,Shapere-book,Dum-PRL-1996}.  As shown in
\cite{Juzeliunas-PRL-2004,Ruseckas-preprint} a non-vanishing effective magnetic
field can arise e.g. if $\Lambda$-type atoms interact with pairs of 
laser fields
that possess a relative orbital angular momentum.  The advantage of this 
scheme
as compared to rotating traps is that it is not limited to
rotationally symmetric configurations. Furthermore in the rotating traps only a
constant effective magnetic field is created \cite{rotating-traps}, whereas
using optical means the effective magnetic field can be controlled and shaped
\cite{Ruseckas-preprint}.
The description of the adiababatic motion of atoms
in terms of gauge potentials has been generalized to  $j+1\to j$
transitions in \cite{Visser-PRA-1998}. The effects of 
gauge potentials on
strongly interacting, bosonic atoms in one-dimensional optical lattices have
been analyzed \cite{Krutizki-PRL-2003}, where it was shown that they lead to
interesting modifications of the Bose-Hubbard model.  An alternative way to
create artificial magnetic fields in lattice gases was recently
suggested employing laser assisted, state-dependent
tunneling \cite{Jaksch,mueller04} or
oscillating potentials with spatial modulations \cite{Sorensen-PRL-2004}.  
In all of these systems the gauge fields have however
$\mathsf{U}(1)$ symmetry, i.e. they are \textit{Abelian}.

As shown by Wilczek and Zee, \textit{non-Abelian} gauge fields can arise
in the adiabatic dynamics of quantum systems with multiple degenerate
eigenstates \cite{Wilczek-PRL-1984}. One of the interesting properties of
non-Abelian gauge potentials is the possibility of magnetic
monopoles.  The presence of
effective magnetic monopole fields in simple quantum systems was 
first pointed out by
Moody, Shapere and Wilczek discussing the adiabatic nuclear rotation in a
diatomic molecule \cite{Moody-PRL-1986}.  
In the present paper we propose an
experimentally realizable scheme which allows to study the motion and the
interaction of neutral quantum gases in non-Abelian gauge fields.  We
show in particular that the coupling of multi-level atoms to spatially varying
laser fields can give rise to such potentials for the atomic 
center-of-mass motion. A necessary condition for this is that the 
atom-laser interaction has degenerate dark eigenstates with
a non-vanishing non-adiabatic coupling.  

\begin{figure}[htb]
\begin{center}
\includegraphics[width=8 true cm]{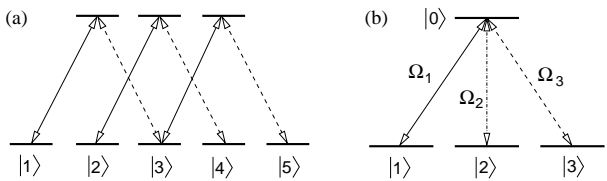}
\end{center}
\caption{\textit{(a)} $j=2 \to j=1$ transition with two degenerate dark states in
the manifolds $\{|1\rangle, |3\rangle, |5\rangle\}$ and
$\{|2\rangle,|4\rangle\}$ which are not coupled by non-adiabatic transitions.
\textit{(b)} Tripod coupling scheme forming two degenerate dark states with
non-adiabatic coupling.}
\label{tripod-scheme}
\end{figure}

Gauge structures in atomic systems with
multiple degenerate dark states have first been discussed by Visser and
Nienhuis \cite{Visser-PRA-1998} considering atoms with a $j+1\to j$ ($j>1$)
transition driven by circularly polarized laser light, as shown in 
  Fig.\ \ref{tripod-scheme}(a).  Since in such a scheme
the dark states are exactly
decoupled, the associated gauge potentials have however again $\mathsf{U}(1)$
symmetry. The simplest system with a non-vanishing adiabatic coupling between
degenerate dark states is the so-called tripod-scheme shown in Fig.\
\ref{tripod-scheme}(b) \cite{Unanyan-OptComm-1998}.  For this scheme the
possibility of non-Abelian topological phases has been predicted analyzing 
fractional adiabatic transfer in \cite{Unanyan-PRA-1999}. Furthermore 
applications to
geometric quantum computation have been put forward and investigated in
\cite{Cirac-Science-2000,Unanyan-PRA-2004}.

After a general discussion of non-Abelian
gauge potentials in the adiabatic motion of atoms in laser fields, we will
introduce the tripod coupling scheme as the simplest system leading to
non-Abelian gauge fields. We then will discuss specific examples. In
particular we will show that using orthogonal laser beams with orbital angular
momentum an effective magnetic field can be generated that has a monopole component.

We start by extending the discussion of Wilczek and Zee
\cite{Wilczek-PRL-1984} and Moody, Shapere and Wilczek \cite{Moody-PRL-1986} to
the adiabatic motion of atoms in stationary laser fields.  For this we consider
atoms with multiple internal states.  
For fixed position $\mathbf{r}$ the internal
Hamiltonian $\hat{H}_{0}(\mathbf{r})$ including the laser interaction can be
diagonalized to give a set of say $N$ dressed states $\left| \chi_n(\mathbf{r})
\right\rangle $ with eigenvalues $\varepsilon _{n}(\mathbf{r})$, where
$n=1,2,\ldots ,N$. The full quantum state of the atom describing both internal
and motional degrees of freedom can then be expanded in terms of the dressed
states according to $|\Phi\rangle =\sum_{n=1}^{N}\Psi _{n}(\mathbf{r}) \left|
\chi_n(\mathbf{r})\right\rangle$.
The $N$-dimensional column vector of wave-functions
$\Psi=(\Psi_1,\Psi_2,\dots,\Psi_N)^\top$ obeys the Schr\"odinger equation
\begin{equation}
i\hbar \frac{\partial}{\partial t}\Psi =\left[\frac{1}{2m}
(-i\hbar\nabla - \mathbf{A})^2 + V\right] \Psi ,
\label{eq:matrix}
\end{equation}
$m$ being the mass of the atoms, and ${V}$ being an external potential that
confines the motion of atoms to a finite region in space.  Here $\mathbf{A}$
and $V$ are $N\times N$ matrices appearing due to the position dependence of
the atomic dressed states:
\begin{eqnarray}
\mathbf{A}_{n,m} &=& i\hbar \langle \chi_n(\mathbf{r})|
\nabla\chi_m(\mathbf{r}) \rangle , \label{eq:A} \\
V_{n,m} &=& \varepsilon _{n}(\mathbf{r})\, \delta_{n,m}
+\langle \chi_n(\mathbf{r})|V(\mathbf{r})|\chi_m(\mathbf{r})\rangle .
\label{eq:V}
\end{eqnarray}
The off-diagonal elements of the matrices $\mathbf{A}$ and $V$ are typically
much smaller than the difference of the dressed atomic energies.  In this case
an adiabatic approximation can be applied which amounts to neglecting the
off-diagonal contributions. This leads to a separation of the dynamics:  
Atoms in
any one of the dressed states evolve according to a separate Hamiltonian with a
$\mathsf{U}(1)$, i.e. Abelian gauge potential.

The adiabatic approximation fails however if there are degenerate (or nearly
degenerate) dressed states. This is the case we are interested in.
Off-diagonal (non-adiabatic) couplings between the degenerate dressed states
can then no longer be ignored.  Suppose the first $q$ atomic dressed states are
degenerate (or nearly degenerate), and these levels are well separated from the
remaining $N-q$. Neglecting transitions to the remaining states, i.e.
projecting the full Hamiltonian to this subspace leads to the Schr\"odinger
equation for the reduced column vector
$\tilde\Psi=\left(\Psi_1,\dots,\Psi_q\right)^\top$
\begin{equation}
i\hbar\frac{\partial}{\partial t}\tilde\Psi=\left[ \frac{1}{2m}
(-i\hbar\nabla -\mathbf{A})^2 + V +\Phi \right]\tilde\Psi
\label{eq:SE-reduced}
\end{equation}
with $\mathbf{A}$ and $V$ being the truncated $q\times q$ matrices.  The
projection of the term $\mathbf{A}^2$ to the $q$ dimensional subspace cannot
entirely be expressed in terms of a truncated matrix $\mathbf{A}$.  This gives
rise to a scalar potential $\Phi$ which is again a $q\times q$ matrix,
\begin{eqnarray}
\Phi _{n,m} &=&\frac{1}{2m}\sum_{l=q+1}^{N}\mathbf{A}_{n,l}\cdot
\mathbf{A}_{l,m}
\label{eq:fi1}  \\
&=&\frac{\hbar ^2}{2m}\left( \langle\nabla\chi _{n}|\nabla\chi_{m}\rangle
+\sum_{k=1}^{q}\langle\chi _{n}|\nabla\chi _{k}\rangle
\langle\chi _{k}|\nabla\chi _{m}\rangle \right)   \nonumber
\end{eqnarray}
with $n,m\in(1,\dots,q)$.  The  reduced 
$q\times q$ matrix $\mathbf{A}$ is called
the Berry connection.

Since the adiabatic states $|\chi_1\rangle \dots |\chi_q\rangle$ are
degenerate, any basis generated by a local unitary transformation
$U(\mathbf{r})$ within the subspace is equivalent. The corresponding local
basis change
\begin{equation}
\tilde\Psi \rightarrow U(\mathbf{r})\tilde\Psi
\label{basis-trafo}
\end{equation}
leads to a transformation of the potentials according to
\begin{eqnarray}
\mathbf{A} &\rightarrow & U(\mathbf{r}) \mathbf{A} U^{\dag}(\mathbf{r})
-i\hbar\left(\nabla U(\mathbf{r})\right) U^{\dag}(\mathbf{r}) ,\\
\Phi &\rightarrow & U(\mathbf{r}) \Phi U^{\dag}(\mathbf{r}) .
\end{eqnarray}
These transformation rules show the gauge character of the potentials
$\mathbf{A}$ and $\Phi$.

The Berry connection or vector potential $\mathbf{A}$ 
is related to a curvature (an effective
``magnetic'' field) $\mathbf{B}$ as:
\begin{eqnarray}
B_i = \frac{1}{2}\epsilon_{ikl} \, F_{kl},\quad
F_{kl} = \partial_k A_l-\partial_l A_k -\frac{i}{\hbar}[A_k,A_l].\label{eq:B}
\end{eqnarray}
Note that the term $\frac{1}{2}\varepsilon_{ikl}[A_k,A_l] = 
(\mathbf{A}\times \mathbf{A})_i$ does not vanish in general,
since the vector components of $\mathbf{A}$ do not necessarily commute. 
In fact this
term reflects the non-Abelian character of the gauge potentials.

The generalized ``magnetic'' field transforms under local
rotations of the degenerate dressed basis (\ref{basis-trafo}) as
\begin{eqnarray}
\mathbf{B} &\rightarrow & U(\mathbf{r})\mathbf{B} U^{\dag}(\mathbf{r}).
\end{eqnarray}
Thus, as expected, $\mathbf{B}$ is a true gauge field.

We will now construct a scheme of laser-atom interactions that leads to a
$\sf{U}(2)$ gauge potential. The 
first requirement is the presence of 
degenerate (or nearly degenerate) dressed states.  Such a condition is
fulfilled e.g.  for the two systems shown in Fig.\ \ref{tripod-scheme}.
Each of
them has two degenerate dark states \cite{Arimondo-review}, i.e. dressed
eigenstates with no component of the excited, radiatively decaying level. Thus
the gauge potentials are $2\times 2$ matrices. In order for them to be truly
non-Abelian, the off-diagonal element $i\hbar \langle\chi_1(\mathbf{r})|\nabla
\chi_2(\mathbf{r})\rangle$ has to be non-zero. One can easily check that this
expression always vanishes for the system discussed in \cite{Visser-PRA-1998}
and shown in Fig.\ \ref{tripod-scheme}(a). It is non-vanishing however for the
so-called tripod scheme shown in Fig.\ \ref{tripod-scheme}(b)
\cite{Unanyan-OptComm-1998}.

The Hamiltonian of the tripod system reads in interaction representation as
\begin{equation}
\hat{H}_0=-\hbar\Bigl(\Omega_1|0\rangle\langle 1|+\Omega_2|0\rangle\langle 2|
+\Omega_3|0\rangle\langle 3|\Bigr)+H.c.,
\end{equation}
Parameterizing the Rabi-frequencies $\Omega_\mu$ with angle and phase variables
according to
\begin{eqnarray}
\Omega_1 & = &\Omega\, \sin\theta\, \cos\phi\, \mathrm{e}^{iS_1},\nonumber \\
\Omega_2 & = &\Omega\, \sin\theta\, \sin\phi\,  \mathrm{e}^{iS_2}, \\
\Omega_3 & = &\Omega\, \cos\theta\, \mathrm{e}^{iS_3}, \nonumber
\end{eqnarray}
where $\Omega =\sqrt{|\Omega_1|^2+|\Omega_2|^2+|\Omega_3|^2}$, the adiabatic
dark states read
\begin{eqnarray}
|D_1\rangle & = &\sin\phi \mathrm{e}^{iS_{31}}|1\rangle
-\cos\phi \mathrm{e}^{iS_{32}}|2\rangle,
\label{eq:D1} \\
|D_2\rangle & = &\cos\theta \cos\phi \mathrm{e}^{iS_{31}}|1\rangle
+\cos\theta\sin\phi \mathrm{e}^{iS_{32}}|2\rangle \nonumber\\
&&  -\sin\theta |3\rangle ,
\label{eq:D2}
\end{eqnarray}
with $S_{ij}=S_i-S_j$. It is now straight-forward to calculate the vector and
scalar gauge potentials. This yields
\begin{eqnarray}
\mathbf{A}_{11} &=& \hbar\left(\cos^2\phi\nabla S_{23}
+ \sin^2\phi\nabla S_{13}\right)\, ,\nonumber \\
\mathbf{A}_{12} &=& \hbar\cos\theta\left(\frac{1}{2}\sin(2\phi)
\nabla S_{12}-i\nabla\phi\right)\, , \label{eq:A-special} \\
\mathbf{A}_{22} &=&\hbar\cos^2\theta\left(\cos^2\phi
\nabla S_{13}+\sin^2\phi\nabla S_{23}\right),\nonumber
\end{eqnarray}
and
\begin{align}
\Phi_{11} & = \frac{\hbar^2}{2m}\sin^2\theta\left(\frac{1}{4}
\sin^2(2\phi)(\nabla S_{12})^2+(\nabla\phi)^2\right),
\nonumber \\
\Phi_{12} & = \frac{\hbar^2}{2m}\sin\theta
\left(\frac{1}{2}\sin(2\phi)\nabla S_{12}
-i\nabla\phi\right)\\
 & \left(\frac{1}{2}\sin(2\theta)(\cos^2\phi\nabla S_{13}
+\sin^2\phi\nabla S_{23})-i\nabla\theta\right), \nonumber\\
\Phi_{22} & = \frac{\hbar^2}{2m}\biggl(\frac{1}{4}\sin^2(2\theta)\left(
\cos^2\phi\nabla S_{13}+\sin^2\phi\nabla S_{23}\right)^2 \nonumber \\
& +(\nabla\theta)^2\biggr).\nonumber
\end{align}
Since the level scheme considered in Fig.\ \ref{tripod-scheme} corresponds to
that of Alkali atoms where $|1\rangle,|2\rangle$, and $|3\rangle$ are Zeeman
components of hyperfine levels, it is natural to assume that the external
trapping potential is diagonal in these states and has the form $V=V_1(
\mathbf{r})|1\rangle\langle 1|+V_2(\mathbf{r})|2\rangle\langle 2|+V_3(
\mathbf{r})|3\rangle\langle 3|$. This still takes into account  the fact that
magnetic, magneto-optical or optical dipole forces can be different in
different Zeeman states.  According to Eq.~(\ref{eq:V}), the external potential
in the adiabatic basis is then given by a $2\times 2$ matrix with elements  $
V_{jk}=\langle D_j|V|D_k\rangle$.  Using the expressions for the dark states
(\ref{eq:D1}) and (\ref{eq:D2}), we arrive at
\begin{eqnarray}
V_{11} &=& V_2\cos^2\phi +V_1\sin^2\phi,\nonumber\\
V_{12}  &=& \frac{1}{2}(V_1-V_2)\cos\theta
\sin(2\phi), \\
V_{22} &=& (V_1\cos^2\phi+V_2\sin^2\phi)\cos^2\theta
+ V_3\sin^2\theta.\nonumber
\end{eqnarray}

At this point it is instructive to consider some specific examples. Let us 
first assume 
that the laser fields coupling levels $|1\rangle$ and $|2\rangle$ 
are co-propagating, and have the same 
frequency and the same orbital angular momentum (if any). In this case their relative phase is fixed 
and can be set $S_{12}=0$. This leads to $S_{13}=S_{23}\equiv S$ and
the expressions for the vector potential simplify to 
\begin{equation}
\mathbf{A}=\hbar\left( 
\begin{array}{cc} 
\nabla S & -i\cos\theta\nabla\phi\, \\ 
i\cos\theta\nabla\phi\, & \cos^2\theta\nabla S 
\end{array} 
\right).
\end{equation}
The components of the $2\times 2$ matrix of the effective magnetic field can 
be easily evaluated and read 
\begin{eqnarray} 
\mathbf{B}_{11} & = & 0, \nonumber\\ 
\mathbf{B}_{12} & = & i\hbar\sin\theta \mathrm{e}^{-iS} 
\nabla\theta \times \nabla\phi \\ 
&& -\hbar\cos\theta \mathrm{e}^{-iS}\nabla S 
\times \nabla\phi (1+\cos^2\theta) , \nonumber\\ 
\mathbf{B}_{22} & = & -2\hbar\cos\theta \sin\theta 
\nabla\theta \times \nabla S. \nonumber
\end{eqnarray} 
One recognizes that a large magnetic field requires large gradients of the 
relative intensities of the fields, parametrized by the angles $\phi$ and 
$\theta$ and a large gradient of the relative phase $S$.  Gradients of $\phi$ 
and $\theta$ on the order of the wavenumber $k$ can be achieved by using 
standing-wave fields.  Large gradients of $S$ can be 
obtained from a running wave $\Omega_3$ orthogonal to the other two or by a 
vortex beam with large orbital angular momentum. 
In this case magnetic fluxes as large as one (in normalized units) 
can be reached.

We now construct a specific field configuration that leads to a
magnetic monopole. For this let us consider two co-propagating and
circularly polarized fields $\Omega_{1,2}$ with opposite orbital angular
momenta $\pm \hbar$ along the propagation axis $z$.  The field $\Omega_3$ propagates in $x$ direction and is linearly polarized
along the $y$-axis: 
\begin{equation}
\Omega_{1,2} = \Omega_0 \frac{\rho}{R}\, \mathrm{e}^{i(k z \mp\varphi)},
\qquad \Omega_3 =\Omega_0 \frac{z}{R}\,  \mathrm{e}^{i k^\prime x}.
\label{eq:Omega-monopole}
\end{equation}
Here $\rho$ is the distance from the $z$-axis and $\varphi$ the azimuthal angle
around this axis. It should be noted that these fields have a vanishing
divergence and obey the Helmholtz equation. This in contrast to the fields
which have been suggested 
in \cite{Zhang-preprint} to create an Abelian magnetic
monopole in a $\Lambda$ system. The total intensity of the laser fields
(\ref{eq:Omega-monopole}) vanishes at a origin, which 
is a singular point. 

The vector potential associated with the
fields can be calculated from Eq.\ (\ref{eq:A-special}). It reads  
\begin{eqnarray}
&&\mathbf{A} =
-\hbar\, \frac{\cos\vartheta}{r\sin\vartheta}\, \hat{\mathrm{e}}_\varphi 
\left(\begin{array}{cc} 0 & 1 \\ 1 & 0 \end{array}\right)+\frac{\hbar}{2}(
k\hat{\mathrm{e}}_z-k^\prime \hat{\mathrm{e}}_x)\times\\
&& \times
\left[(1+\cos^2\vartheta) 
\left(\begin{array}{cc} 1 & 0 \\ 0 & 1 \end{array}\right)
+(1-\cos^2\vartheta)
\left(\begin{array}{cc} 1 & 0 \\ 0 & -1 \end{array}\right)\right]. \nonumber
\end{eqnarray} 
The first term proportional to $\sigma_x$ corresponds
to a magnetic monopole of strength one at the origin.
This is easily seen by calculating the magnetic field
\begin{equation}
\mathbf{B}= \frac{\hbar}{r^2}\, \hat{\mathrm e}_r\,
\left(\begin{array}{cc} 0 & 1 \\ 1 & 0 \end{array}\right)
 + \cdots\, .
\end{equation}
The dots indicate non-monopole field contributions proportional to $\sigma_z,
\sigma_y$ and the unity matrix.

In the present paper we have shown that the adiabatic motion of multi-level
atoms interacting with spatially varying laser fields in the tripod-coupling
configuration can lead to $\mathsf{U}(2)$ non-Abelian gauge potentials. The
system can easily be generalized to effective $\mathsf{U}(n)$, $n>2$,
 gauge structures using atomic configurations with more than three
laser fields coupling to a common excited state.  The strength of the effective
magnetic fields can be large if standing wave configurations or light beams
with large orbital angular momentum are used.  As a specific example we
have identified a configuration of laser fields which leads to a magnetic
monopole.  

Our approach is complementary to the recent proposal of
Osterloh \textit{et al.} \cite{Osterloh}, who suggested the generation of
effective non-Abelian fields in lattice gases. For this they employed a
state-dependent manipulation of tunneling amplitudes by lasers.
These proposals make the study of interacting degenerate Bose or
Fermi gases in non-Abelian gauge fields experimentally
feasible.


\begin{acknowledgments}
M.F. would like to thank R.G. Unanyan for discussions. 
This work has been supported by the Alexander-von-Humboldt 
foundation. 
J.R.\ has been supported by the EU through the Marie-Curie Trainingssite at 
the TU Kaiserslautern. P.\"O.\ wishes to acknowledge the 
Royal Society of Edinburg
for support.
\end{acknowledgments}


\end{document}